\documentstyle[aps,prd,preprint,floats]{revtex}
\input epsf
\tightenlines

\newcommand{\beq}{\begin{equation}}
\newcommand{\eeq}{\end{equation}}

\def\lap{\lower.5ex\hbox{$\; \buildrel < \over \sim \;$}}
\def\gap{\lower.5ex\hbox{$\; \buildrel > \over \sim \;$}}

\def\L{\Lambda}
\def\rV{\rho_v}
\def\rL{\rho_\L}

\begin{document}

\title{Anthropic predictions: the case of the cosmological constant}

\author{Alexander Vilenkin}

\address{ Institute of Cosmology, Department of Physics and Astronomy,\\ 
Tufts University, Medford, MA 02155, USA}

\maketitle

\begin{abstract}

Anthropic models can give testable predictions, which can be confirmed
or falsified at a specified confidence level. This is illustrated
using the successful prediction of the cosmological constant as an
example. The history and the nature of the prediction are reviewed.
Inclusion of other variable parameters and implications for particle
physics are briefly discussed.

\end{abstract}


\section{Introduction}

The parameters we call constants of Nature may in fact be stochastic
variables taking different values in different parts of the universe.
The observed values of these parameters are then determined by chance
and by anthropic selection.  It has been argued, at least for some of the
constants, that only a narrow range of their values is consistent with
the existence of life \cite{Carter,Carr,Barrow,TR,Hogan}.

These arguments have not been taken very seriously and have often been
ridiculed as handwaving and unpredictive.  For one thing, the
anthropic worldview assumes some sort of a ``multiverse'' ensemble,
consisting of multiple universes or distant regions of the same
universe, with constants of Nature varying from one member of this
ensemble to another.  Quantitative results cannot be obtained without
a theory of the multiverse. Another criticism is that the anthropic
approach does not make testable predictions; thus it is not
falsifiable, and therefore not scientific.

While both of these criticisms had some force a couple of decades ago,
much progress has been made since then, and the situation is now
completely different. The first criticism no longer applies, because
we now do have a theory of the multiverse.  It is the theory of
inflation. A remarkable feature of inflation is that, generically, it
never ends completely.  The end of inflation is a stochastic process;
it occurs at different times in different parts of the universe, and
at any time there are regions which are still inflating
\cite{AV83,Linde86}.  If some ``constants'' of Nature are related to
dynamical fields and are allowed to vary, they are necessarily
randomized by quantum fluctuations during inflation and take different
values in different parts of the universe.  Thus, inflationary
cosmology gives a specific realization of the multiverse ensemble, and
makes it essentially inevitable. (For a review see, e.g.,
\cite{Lindebook}.)

In this paper I am going to address the second criticism, that
anthropic arguments are unpredictive. I will try to dispel this notion
and outline how anthropic models can be used to make quantitative
predictions. These predictions are of a statistical nature, but they
still allow models to be confirmed or falsified at a specified
confidence level. I will focus on the case of the cosmological
constant, whose nonzero value was predicted anthropically well before
it was observed. This case is of great interest in its own right and
is well suited to illustrate the issues associated with anthropic
predictions.

\section{Anthropic bounds {\it vs.} anthropic predictions} 

For terminological clarity, it is important to distinguish between
anthropic bounds and anthropic predictions. Suppose there is some
parameter $X$, which varies from one place in the universe to
another. Suppose further that the value of $X$ affects the chances for
intelligent observers to evolve, and that the evolution of 
observers is possible only if $X$ is within some interval
\beq
X_{min}<X<X_{max}.
\label{range} 
\eeq
Clearly, values of $X$ outside the interval (\ref{range}) are not
going to be observed, because such values are inconsistent with the
existence of observers. This statement is often called ``the anthropic
principle''.

Although anthropic bounds, like Eq.(\ref{range}), can have considerable
explanatory power, they can hardly be regarded as predictions: they
are guarranteed to be right.  And the ``anthropic principle'', as
stated above, hardly deserves to be called a principle: it is
trivially true. This is not to say, however, that anthropic arguments
cannot yield testable predictions.

Suppose we want to test a theory according to which the parameter $X$
varies from one part of the universe to another.\footnote{I assume for
simplicity that $X$ is variable only in space, but not in time.} Then,
instead of looking for the extreme values $X_{min}$ and $X_{max}$ that
make observers impossible, we can try to predict what values of $X$
will be measured by typical observers. In other words, we can make
statistical predictions, assigning probabilities $P(X)$ to different
values of $X$. [$P(X)$ is the probability that an observer randomly
picked in the universe will measure a given value of $X$.] If any
principle needs to be invoked here, it is what I called ``the
principle of mediocrity'' \cite{AV95} -- the assumption that we are
typical among the observers in the universe. Quantitatively, this can
be expressed as the expectation that we should find ourselves, say,
within the 95\% range of the distribution. This can be regarded as a
prediction at a 95\% confidence level. If instead we measure a value
outside the expected range, this should be regarded as evidence
against the theory.

\section{The cosmological constant problem}

The cosmological constant is (up to a factor) the energy density of
the vacuum, $\rho_v$. Below, I do not distinguish between the two and
use the terms ``cosmological constant'' and ``vacuum energy density''
interchangeably. By Einstein's mass-energy relation, the energy
density is simply related to the mass density, and I will often
express $\rV$ in units of g/cm$^3$.  

The gravitational properties of the vacuum are rather unusual: for
positive $\rho_v$, its gravitational force is repulsive. This can be
traced to the fact that, according to Einstein's General Relativity,
the force of gravity is determined not solely by the energy (mass)
density $\rho$, but rather by the combination $(\rho+3P)$, where $P$
is the pressure. In ordinary astrophysical objects, like stars or
galaxies, pressure is much smaller than the energy density, $P\ll
\rho$, and its contribution to gravity can be neglected. But in the
case of vacuum, the pressure is equal and opposite to
$\rV$,\footnote{Since the vacuum energy is proportional to the volume
$V$ it occupies, $E=\rho_v V$, the pressure is $P_v=-d E/d
V=-\rho_v$.}
\beq
P_v =-\rho_v,
\label{Pv}
\eeq
so that $\rho_v+3P_v=-2\rho_v$. Pressure not only contributes
significantly to the gravitational force produced by the mass, it also
changes its sign.

The cosmological constant was introduced by Einstein in his 1917 paper
\cite{Einstein}, where he applied the newly developed theory of General 
Relativity to the universe as a whole. Einstein believed that
the universe was static, but to his dismay he found that the theory had
no static cosmological solutions. He concluded that the theory had to
be modified and introduced the cosmological term, which amounted to
endowing the vacuum with a positive energy density. The magnitude of
$\rho_v$ was chosen so that its repulsive gravity exactly balanced the
attractive gravity of matter, resulting in a static world. More than a
decade later, after Hubble's discovery of the expansion of the
universe, Einstein abandoned the cosmological constant, calling it the
greatest blunder of his life. But once the Genie was out of the
bottle, it was not so easy to put it back.

Even if we do not introduce the vacuum energy ``by hand'',
fluctuations of quantum fields, like the electromagnetic field, would
still make this energy nonzero. Adding up the energies of quantum
fluctuations with shorter and shorter wavelengths gives a formally
infinite answer for $\rho_v$. The sum has to be cut off at the Planck
length, $l_P\sim 10^{-33}$ cm, where quantum gravity effects become
important and the usual concepts of space and time no longer
apply. This gives a finite, but absurdly large value, $\rho_v\sim
10^{94}$ g/cm$^3$. A cosmological constant of this magnitude would
cause the universe to expand with a stupendous acceleration.  If
indeed our vacuum has energy, it should be at least 120 orders of
magnitude smaller in order to be consistent with observations. In
supersymmetric theories, the contributions of different fields
partially cancel, and the discrepancy can be reduced to 60 orders of
magnitude.  This discrepancy between the expected and observed values
of $\rho_v$ is called the cosmological constant problem. It is one of
the most intriguing mysteries that we are now facing in theoretical
physics.

\section{The anthropic bound}

A natural resolution to the cosmological constant problem is obtained
in models where $\rho_v$ is a random variable.  The idea is to
introduce a dynamical dark energy component $X$ whose energy density
$\rho_X$ varies from place to place, due to stochastic processes that
occured in the early universe.  A possible model for $\rho_X$ is a
scalar field with a very flat potential \cite{Linde86',GV00}, such
that the field is driven to its minimum on an extremely long
timescale, much longer than the present age of the universe. Another
possibility is a discrete set of vacuum states. Transitions between
different states can then occur through nucleation and expansion of
bubbles bounded by domain walls \cite{Teitelboim,Abbott}.  The effective
cosmological constant is given by $\rho_v=\rho_\Lambda+\rho_X$, where
$\rL$ is the constant vacuum energy density, which may be as large as
(+ or -)$10^{94}$ g/cm$^3$. The cosmological constant problem now takes a
different form: the puzzle is why we happen to live in a region where
$\rL$ is nearly cancelled by $\rho_X$.

The key observation, due to Weinberg \cite{Weinberg87} (see also
\cite{Barrow,Linde86',Davies}) is that the cosmological constant can 
have a dramatic effect on the formation of structure in the universe.
The observed structures - stars, galaxies, and galaxy clusters -
evolved from small initial inhomogeneities, which grew over eons of
cosmic time by gravitationally attracting matter from surrounding
regions. As the universe expands, matter is diluted, so its density
goes down as
\beq
\rho_M = (1+z)^3\rho_{M0}, 
\label{rhoM}
\eeq
where $\rho_{M0}$ is the present matter density and $z$ is the
redshift.\footnote{The redshift $z$ is defined so that $(1+z)$ is the
expansion factor of the universe between a given epoch and the present
(earlier times correspond to larger redshifts).} At the same time, the
density contrast $\sigma\equiv\delta\rho/\rho$ between overdense and underdense
regions keeps growing. Gravitationally bound objects form where
$\sigma \sim 1$. The first stars form in relatively small
matter clumps of mass $\sim 10^6 M_\odot$. The clumps then merge into
larger and larger objects, leading to the formation of giant galaxies
like ours and of galaxy clusters.

How is this picture modified in the presence of a cosmological
constant?  At early times, when the density of matter is high,
$\rho_M\gg\rho_v$, the vacuum energy has very little effect on
structure formation. But as the universe expands and the matter
density decreases, the vacuum density $\rho_v$ remains
constant and eventually becomes greater than $\rho_M$.
At this point the character of cosmic expansion changes.
Prior to vacuum domination, the expansion is slowed down by gravity,
but afterwards it begins to accelerate, due to the repulsive gravity
of the vacuum. Weinberg showed that the growth of density
inhomogeneities effectively stops at that epoch. If no structures were
formed at earlier times, then none will ever be formed.

It seems reasonable to assume that the existence of stars is a
necessary prerequisite for the evolution of observers. We also need to
require that the stars belong to sufficiently large bound objects -
galaxies - so that their gravity is strong enough to retain the heavy
elements dispersed in supernova explosions. These elements are
necessary for the formation of planets and of observers.  An anthropic
bound on the vacuum energy can then be obtained by requiring that
$\rho_v$ does not dominate before the redshift $z_{max}$ when the
earliest galaxies are formed. With the aid of Eq.(\ref{rhoM}), this
gives
\beq
\rho_v \lesssim (1+z_{max})^3 \rho_{M0}.
\label{bound}
\eeq 
The most distant galaxies observed at the time when Weinberg
wrote his paper had redshifts $z\sim 4.5$. Assuming that $z_{max}\sim 4.5$,
Eq.(\ref{bound}) yields the bound $\rho_v\lesssim 170\rho_{M0}$. A
more careful analysis by Weinberg showed that in order to prevent
structure formation, $\rho_v$ needs to be 3 times greater than
suggested by Eq.(\ref{bound}); hence, a more accurate bound is
\cite{Weinberg87} 
\beq 
\rho_v\lesssim 500 \rho_{M0}.
\label{Wbound}
\eeq
Of course, observation of galaxies at $z\sim 4.5$ means only that
$z_{max}\gtrsim 4.5$, and Weinberg referred to (\ref{Wbound}) as ``a
lower bound on the anthropic upper bound on $\rV$.''
At present, galaxies are observed at considerably higher redshifts, up to
$z\sim 10$. The corresponding bound on $\rho_v$ would be
\beq
\rho_v\lesssim 4000 \rho_{M0}.
\eeq

For negative values of $\rho_v$, the vacuum gravity is attractive, and
vacuum domination leads to a rapid recollapse of the universe. An
anthropic lower bound on $\rho_v$ can be obtained in this case by
requiring that the universe does not recollapse before life had a
chance to develop \cite{Barrow,KL}. Assuming that the timescale for
life evolution is comparable to the present cosmic time, one finds
$\rho_v\gtrsim -\rho_{M0}$.\footnote{An important distinction between
positive and negative values of $\rV$ is that for $\rV >0$, galaxies
that formed prior to vacuum domination can survive indefinitely in
the vacuum-dominated universe.}

The anthropic bounds are narrower, by many orders of magnitude, than the
particle physics estimates for $\rV$. Moreover, as Weinberg noted,
there is a prediction implicit in these bounds. He wrote
\cite{Weinberg89}: ``... if it is the anthropic principle that accounts 
for the smallness of the cosmological constant, then we would expect a
vacuum energy density $\rho_v\sim (10-100)\rho_{M0}$, because there is
no anthropic reason for it to be any smaller.''

One has to admit, however, that the anthropic bounds fall short of the
observational bound, $(\rho_v)_{obs}\lesssim 4\rho_{M0}$, by a few
orders of magnitude. If all the values in the anthropically allowed
range were equally probable, an additional fine-tuning by a factor of
$100-1000$ would still be needed.

\section{Anthropic predictions}

The anthropic bound (\ref{bound}) specifies the value of $\rho_v$
which makes galaxy formation barely possible.  However, if $\rho_v$
varies in space, then most of the galaxies will not be in regions
characterized by these marginal values, but rather in regions where
$\rho_v$ dominates after a substantial fraction of matter had already
clustered into galaxies.

To make this quantitative, we define the probability distribution
${\cal P}(\rV)d\rV$ as being proportional to the number of observers
in the universe who will measure $\rV$ in the interval $d\rV$. This
distribution can be represented as a product \cite{AV95}
\beq
{\cal P}(\rV)d\rV=n_{obs}(\rV){\cal P}_{prior}(\rV) d\rV.
\label{dP}
\eeq
Here, ${\cal P}_{prior}(\rV)d\rV$ is the prior distribution, which is
proportional to the volume of those parts of the universe where
$\rho_v$ takes values in the interval $d\rV$, and $n_{obs}(\rV)$ is
the number of observers that are going to evolve per unit volume.
The distribution (\ref{dP}) gives the probability that a
randomly selected observer is located in a region where the effective
cosmological constant is in the interval $d\rho_v$.  

Of course, we have no idea how to calculate $n_{obs}$, but what comes
to the rescue is the fact that the value of $\rV$ does not directly
affect the physics and chemistry of life. As a rough approximation, we
can then assume that $n_{obs}(\rV)$ is simply proportional to the
fraction of matter $f$ clustered in giant galaxies like ours
(with mass $M\gtrsim M_G = 10^{12}M_{\odot}$),
\beq
n_{obs}(\rV)\propto f(M_G,\rV).
\label{nobs}
\eeq
The idea is that there is a certain number of stars per unit mass in a
galaxy and certain number of observers per star. The choice of the
galactic mass $M_G$ is an important issue; I will comment on it in
next section.

The calculation of the prior distribution ${\cal P}_{prior}(\rV)$
requires a particle physics model which allows $\rV$ to vary and a
cosmological ``multiverse'' model that would generate an ensemble of
sub-universes with different values of $\rV$. An example of a suitable
particle theory is the superstring theory, which appears to admit an
incredibly large number of vacua (possibly as large as $10^{1000}$
\cite{Bousso,Douglas,Susskind}) characterized by different 
values of particle masses, couplings, and other parameters, including
the cosmological constant. When this is combined with the cosmic
inflation scenario, one finds that bubbles of different vacua
copiously nucleate and expand during inflation, producing
exponentially large regions with all possible values of $\rV$. Given a
particle physics model and a model of inflation, one can in principle
calculate ${\cal P}_{prior}(\rV)$. Examples of calculation for
specific models have been given in
\cite{GV00,GV01,GLindeV}.\footnote{There are still some unresolved issues
regarding the calculation of ${\cal P}_{prior}$ for models with a discrete spectrum of variable ``constants''. For a discussion see \cite{LLM,GBL,markers}.} 
Needless to say, the details of the fundamental theory and of the
inflationary dynamics are too uncertain for a definitive calculation
of ${\cal P}_{prior}$. We shall instead rely on the following general
argument \cite{AV96,Weinberg96}.

Suppose some parameter $X$ varies in the range $\Delta X$ and is
characterized by a prior distribution ${\cal P}_{prior}(X)$. Suppose
further that $X$ affects the number of observers in such a way that
this number is non-negligible only in a very narrow range $\Delta
X_{obs}\ll \Delta X$. Then one can expect that the function ${\cal
P}_{prior}(X)$ with a large characteristic range of variation should
be very nearly a constant in the tiny interval $\Delta X_{obs}$. In
the case of $\rho_v$, the range $\Delta\rV$ is set by the Planck scale
or by the supersymmetry breaking scale, and we have $(\Delta
\rV)_{obs}/\Delta\rV \sim 10^{-60} -10^{-120}$. Hence, we 
expect
\beq
{\cal P}_{prior}(\rV)\approx {\rm const}.
\label{prior}
\eeq
I emphasize that the assumption here is that the value $\rV=0$ is not
in any way special, as far as the fundamental theory is concerned, and
is, therefore, not a singular point of ${\cal P}_{prior}(\rV)$.

Combining Eqs.(\ref{dP}),(\ref{nobs}),(\ref{prior}), we obtain
\beq
{\cal P}(\rV)\propto f(M_G,\rV).
\label{Pf}
\eeq

In Ref.\cite{AV95}, where I first introduced the anthropic probability
distributions of the form (\ref{dP}), I did not attempt a detailed
calculation of the distribution for $\rV$, resorting instead to a
rough estimate. If we denote by $z_G$ the redshift at the epoch of
galaxy formation, then most of the galaxies should be in regions where
the vacuum energy dominates at $z_v\lesssim z_G$. Regions with $z_v\gg
z_G$ will have very few galaxies, while regions with $z_v\ll z_G$ will
be rare, simply because they correspond to a very narrow range of
$\rV$ near zero. Hence, we expect a typical galaxy to be
located in a region where 
\beq
z_v\sim z_G.  
\label{coincidence}
\eeq
The expected value of $\rV$ is then 
\beq
\rV\sim (1+z_G)^3\rho_{M0}.  
\label{zG}
\eeq
The choice of the galaxy formation epoch $z_G$ is related to the
choice of the galactic mass $M_G$ in (\ref{nobs}). I used $z_G\sim
1$, obtaining $\rV\sim 8\rho_{M0}$.

A similar approach was later developed by Efstathiou
\cite{Efstathiou}. The main difference is that he calculated the
fraction of clustered matter $f$ at the time corresponding to the
observed value of the microwave background temperature, $T_0=2.73~K$,
while my suggestion was to use the asymptotic value of $f$ at $t\to
\infty$.  The two approaches correspond to different choices of the
reference class of observers among whom we expect to be
typical. Efstathiou's choice includes (roughly) only observers that
have evolved until present, while my choice is to include all
observers throughout the history of the universe. If we are truly
typical, and live at the time when most observers live, the two
methods should give similar results.  Indeed, one finds that the
probability distributions calculated by these methods are nearly
identical \cite{Levon}.\footnote{The original calculation by
Efstathiou gave a different result, but that calculation contained an
error, which was later pointed out by Weinberg \cite{Weinberg96}.}

\section{Comparison with observations}

Despite a number of observational hints that the cosmological constant
might be nonzero (see, e.g., \cite{Fukugita}), its discovery still
came as a great surprise to most physicists and
astronomers. Observations of distant supernovae by two independent
groups in 1997-98 provided strong evidence that the expansion of the
universe is accelerating \cite{Supernova}. The simplest interpretation
of the data was in terms of a cosmological constant with $\rV\sim
2.3\rho_{M0}$. Further evidence came from the cosmic microwave
background and galaxy clustering observations, and by now the case for
the cosmological constant is very strong.

The discovery of the cosmological constant was particularly shocking
to particle physicists who almost universally believed that it should
be equal to zero. They assumed that something so small could only be
zero and searched for a new symmetry principle or a dynamical
adjustment mechanism that would force $\rV$ to vanish. The
observed value of $\rV$ brought yet another
puzzle. The matter density $\rho_M$ and the vacuum energy density
$\rho_v$ scale very differently with the expansion of the universe. In
the early universe the matter density dominates, while in the
asymptotic future it becomes negligible. There is only one epoch in
the history of the universe when $\rho_M\sim \rV$. It is difficult to
understand why we happen to live in this very special epoch. This is
the so-called cosmic coincidence problem.

The coincidence is easily understood in the framework of the anthropic
approach \cite{GLV,Bludman}. The galaxy formation epoch, $z_G\sim
1-3$, is close to the present cosmic time, and the anthropic
model predicts that the vacuum domination should begin at $z\sim
z_G$ [see Eq.~(\ref{coincidence})]. This explains the coincidence.

\begin{figure}
\centerline{\epsfxsize = 0.5\hsize \epsfbox{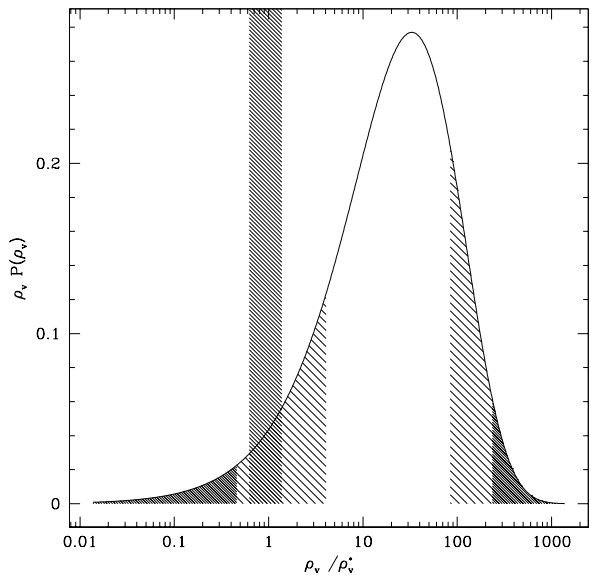}}
\vskip 0.5 truecm
\caption{\label{fig1}
The logarithmic probability distribution $d{\cal P}/d(\log\rV)$. 
The lightly and densely shaded areas are the
regions excluded at 68\% and 95\% level, respectively. The uncertainty
in the observed value $\rV^*$ is indicated by the vertical strip.
}
\end{figure}

The probability distribution for $\rho_v$ based on Eq.(\ref{Pf}) was
extensively analyzed in \cite{MSW}. The distribution depends on the
amplitude of galactic-scale density perturbations, $\sigma$, which can
be specified at some suitably selected epoch (e.g., the epoch of
recombination).  Until recently, significant uncertainties in this
quantity complicated the comparison of anthropic predictions with the
data \cite{MSW,GLindeV}. These uncertainties appear now to have been
mostly resolved \cite{Max}. In Fig.~\ref{fig1} we plot, following
\cite{PTV2}, the resulting probability distribution per logarithmic
interval of $\rV$. Only positive values of $\rV$ are considered, so
this can be regarded as a conditional distribution, given that
$\rV>0$.  On the horizontal axis, $\rV$ is plotted in units of the
observed vacuum energy density, $\rV^* = 7\times 10^{-30}$
g/cm$^3$. The 68\% and 95\% ranges of the distribution are indicated
by light and dark shading, respectively.

We note that the confidence level ranges in Fig.~\ref{fig1} are rather broad.
This corresponds to a genuine large variance in the cosmic
distribution of $\rV$. The median value of the distribution is about
20 times greater than the observed value. But still, the observed
value $\rV^*$ falls well within the range of anthropic prediction at
95\% confidence level.

At this point, I would like to comment on two important assumptions
that went into the successful prediction of the observed value of
$\rV$. First, we assumed a flat prior probability distribution
(\ref{prior}). Analysis of specific models shows that this assumption
is indeed valid in a wide class of models, but it is not as
automatic as one might expect
\cite{GV00,Weinbergcomment,GV01,Predictions}. In particular, it is not clear
that it is applicable to the superstring-inspired models of the type
discussed in \cite{Bousso,Douglas,Susskind} (more on this in Section
VIII). 

Second, we used the value of $M_G = 10^{12} M_\odot$ for the galactic
mass in (\ref{Pf}). This amounts to assuming that most observers live
in giant galaxies like our Milky Way. We know from observations that
some galaxies existed already at $z=10$, and the theory predicts that
some dwarf galaxies and dense central parts of giant galaxies could
form as early as $z=20$. If observers were as likely to evolve in
early galaxies as in late ones, the value of $\rV$ indicated by
Eq.(\ref{zG}) would be far greater than observed. Clearly, the
agreement is much better if we assume that the conditions for
civilizations to emerge arise mainly in galaxies which form at lower
redshifts, $z_G\sim 1$.

Following \cite{Predictions}, I will now point to some directions
along which the choice of $z_G\sim 1$ may be justified. As already
mentioned, one problem with dwarf galaxies is that their mass may be
too small to retain the heavy elements dispersed in supernova
explosions. Numerical simulations suggest that the fraction of heavy
elements retained is $\sim 30\%$ for a $10^9 M_\odot$ galaxy and is
negligible for much smaller galaxies \cite{MacLow}. Hence, we have to
require that the structure formation hierarchy evolves up to
mass scales $\sim 10^9 M_\odot$ or higher prior to the vacuum energy
domination. This gives the condition $z_G\lesssim 3$, but falls short
of explaining $z_G\sim 1$.

Another point to note is that smaller galaxies, formed at earlier
times, have a higher density of matter. This may increase the danger
of nearby supernova explosions and the rate of near encounters with
stars, large molecular clouds, or dark matter clumps. Gravitational
perturbations of planetary systems in such encounters could send a
rain of comets from the Oort-type clouds towards the inner planets,
causing mass extinctions. 

Our own Galaxy has definitely passed the test for the evolution of
observers, and the principle of mediocrity suggests that most
observers may live in galaxies of this type.  Our Milky Way is a giant
spiral galaxy. The dense central parts of such galaxies were formed at
a high redshift $z\gtrsim 5$, but their discs were assembled at
$z\lesssim 1$ \cite{Abraham}. Our Sun is located in the disc, and if
this situation is typical, then the relevant epoch to use in
Eq.(\ref{zG}) is the epoch $z_G\sim 1$ associated with the formation
of discs of giant galaxies.

These remarks may or may not be on the right track, but if the
observed value of $\rV$ is due to anthropic selection, then, for one
reason or another, the evolution of intelligent life should require
conditions which are found mainly in giant galaxies, which completed
their formation at $z\sim 1$. This is a prediction of the anthropic
approach. It will be subject to test when our understanding of
galactic evolution and of the conditions necessary to sustain
habitable planetary systems will reach an adequate level -- hopefully
in not so distant future.

\section{Predictions for the equation of state}

A generic prediction of anthropic models for the vacuum energy is that
the vacuum equation of state (\ref{Pv}) should hold with a very high
accuracy \cite{Predictions}. In models of discrete vacua, this
equation of state is guaranteed by the fact that in each vacuum the
energy density is a constant and can only change by nucleation of
bubbles. If $\rho_X$ is a scalar field potential, it must satisfy the
slow-roll condition -- that the field should change slowly on the time
scale of the present age of the universe. The slow-roll condition is
likely to be satisfied by excess, by many orders of
magnitude. Although it is possible to adjust the potential so that it
is only marginally satisfied, it is satisfied by a very wide margin in
generic models. This implies the equation of state (\ref{Pv}).

There is also a related prediction, which is not likely to be tested
anytime soon.  In anthropic models, $\rV$ can take both positive and
negative values, so the observed positive dark energy will eventually
start decreasing and will turn negative, and our part of the universe
will recollapse to a big crunch.  Since the evolution of $\rV$ is
expected to be very slow on the present Hubble scale, we do not expect
this to happen sooner than in a trillion years from now
\cite{Predictions}.

It should be noted that the situation may be different in more
complicated models, involving more than one scalar field. It has been
shown in \cite{GLindeV} that the equation of state in such models may
significantly deviate from (\ref{Pv}), and the recollapse may occur on
a timescale comparable to the lifetime of the Sun. Observational tests
allowing to distinguish between the two types of models have been
discussed in \cite{Thomas,KK,GPVa}. Recent observations yield
\cite{Max} $P_v /\rV=-1\pm 0.1$, consistent with the simplest models.

\section{Implications for particle physics}

Anthropic models for the cosmological constant have nontrivial
implications for particle physics. Scalar field models require the
existence of fields with extremely flat potentials. Models with a
discrete set of vacua require that the spectrum of values of $\rV$
should be very dense, so that there are many such values in the small
anthropically allowed range.  This points to the existence of very
small parameters that are absent in familiar particle physics
models. Some ideas on how such small parameters could arise have been
suggested in \cite{GV00,Weinbergcomment,Feng,Banks,Donoghue,Dvali}.

A different possibility, which has now attracted much attention, is
inspired by superstring theory. This theory presumably has an enormous
number of different vacua, scattered over a vast ``string theory
landscape''. The spectrum of $\rV$ (and of other particle physics
constants) can then be very dense without any small parameters, due to
the sheer number of vacua \cite{Bousso,Douglas,Susskind}. This
picture, however, entails a potential problem.  Vacua with close
values of $\rV$ are not expected to be close to one another in the
``landscape'', and there seems to be no reason to expect that they
will be chosen with equal probability by the inflationary
dynamics. Hence, we can no longer argue that the prior probability
distribution is flat. In fact, since inflation is characterized by an
exponential expansion of the universe, and the expansion rate is
different in different parts of the landscape, the probabilities for
well separated vacua are likely to differ by large exponential
factors. If indeed the prior distribution is very different from flat,
this may destroy the successful anthropic prediction for $\rV$. This
issue requires further study, and I am sure we are going to hear more
about it.

\section{Including other variables}

If the cosmological constant is variable, then it is natural to expect
that some other ``constants'' could vary as well, and it has been
argued that including other variables may drastically modify the
anthropic prediction for $\rV$ \cite{TR,Aguirre,BDG}. The idea is that
the adverse effect on the evolution of observers due to a change in one
variable may be compensated by an appropriate change in another
variable. As a result the peak of the distribution may drift into a
totally different area of the parameter space.  While this is a
legitimate concern, specific models with more than one variable that
have been analyzed so far suggest that the anthropic prediction for
$\rV$ is rather robust.

Suppose, for example, that $\rV$ and the primordial density contrast
$\sigma$ (specified at recombination) are both allowed to vary. Then
we are interested in the joint distribution 
\beq
{\cal P}(\rho_v,\sigma)d\rho_v d\sigma.
\label{Psigma}
\eeq
Using the same assumptions as in Section V\footnote{The assumption
that the number of observers is simply proportional to the fraction of
matter clustered into galaxies may not give a good approximation in
regions where $\sigma$ is very large. In such regions, galaxies form
early and are very dense, so chances for life to evolve may be
reduced. A more accurate calculation should await better estimates for
the density of habitable stellar systems.} and introducing a new
variable $y=\rV/\sigma^3$, one finds \cite{Predictions} that this
distribution factorizes to the form\footnote{Note that
there is no reason to expect the prior distribution for $\sigma$ to be
flat. The amplitude of density perturbations is related to the
dynamics of the inflaton field that drives inflation and is therefore
strongly correlated with the amount of inflationary expansion. Hence,
we expect ${\cal P}_{prior}$ to be a nontrivial function of
$\sigma$. In fact, it follows from (\ref{Py}) that ${\cal
P}_{prior}(\sigma)$ should decay at least as fast as $\sigma^{-3}$ in
order for the distribution to be integrable \cite{GLV}.}
\beq
\sigma^3 {\cal P}_{prior}(\sigma)d\sigma\cdot f(y)dy,
\label{Py}
\eeq
where, $f(y)$ is the fraction of matter clustered in galaxies (which
depends only on the combination $\rV/\sigma^3=y$). 

After integration over $\sigma$, we obtain essentially the same
distribution as before, but for a new variable $y$. The prediction now
is not for a particular value of $\rV$, but for a relation between
$\rV$ and $\sigma$. Comparison of the predicted and observed values of
$y$ is given by the same graph as in Fig.~1, with a suitable rescaling
of the horizontal axis. As before, the 95\% confidence level
prediction is in agreement with the data.

Another example is a model where the neutrino masses are assumed to be
anthropic variables. Neutrinos are elusive light particles, which
interact very weakly and whose masses are not precisely known. The
current astrophysical upper bound on the neutrino mass is $m_\nu \lesssim
0.5$ eV \cite{Max}, and the lower bound from the neutrino oscillation
data is $m_\nu\gtrsim 0.05$ eV \cite{Bahcall}. (Here and below $m_\nu$
denotes the sum of the three neutrino masses.) It has been suggested
in \cite{PTV1} that small values of the neutrino masses may be due to
anthropic selection. A small increase of $m_\nu$ can have a large
effect on galaxy formation. Neutrinos stream out of overdense regions,
slowing the growth of density perturbations. The fraction of mass that
neutrinos contribute to the total density of the universe is
proportional to $m_\nu$. Thus, perturbations will grow slower, and
there will be fewer galaxies, in regions with larger values of $m_\nu$.
A calculation along the same lines as in Section V yields a
prediction 0.07 eV $< m_\nu <$ 5.7 eV at 95\% confidence level.

In Ref. \cite{PTV2} this model was extended, allowing both $m_\nu$ and
$\rV$ to be anthropic variables. The resulting probability
distribution ${\cal P}(\rV,m_\nu)$ is concentrated in a localized
region of the parameter space. Its peak is not far off from the peaks
of the individual distributions for $\rV$ and $m_\nu$. In fact,
inclusion of $m_\nu$ somewhat improves the agreement of the prediction
for $\rV$ with the data.

The parameters $\rV$, $\sigma$ and $m_\nu$ share the property that
they do not directly affect life processes. Other parameters of this
sort include the mass of dark matter particles and of baryons per
photon. The effects of varying these parameters have been discussed in
\cite{TR,Aguirre}. In particular, Aguirre \cite{Aguirre} argued that 
values of the baryon to photon ratio much higher than the observed may
be anthropically favored. What he showed, in fact, is that this
proposition cannot at present be excluded. This is an interesting
issue and certainly deserves further study. Extensions to parameters
like the electron mass or charge, which do affect life processes, is
on a much shakier ground. Untill these processes are much better
understood, one will have to resort to qualitative arguments, as in
\cite{Carter,Carr,Barrow,Hogan}.

\section{Concluding remarks}

The case of the cosmological constant demonstrates that anthropic
models can be subjected to observational tests and can be confirmed or
ruled out at a specified confidence level. It also illustrates the
limitations and difficulties of anthropic predictions.
 
The situation we are accustomed to in physics is that the agreement
between theory and observations steadily improves, as the
theoretical calculations are refined and the accuracy of measurements
increases. Not so in anthropic models. Here, predictions are in the form
of probability distributions, having an intrinsic variance which
cannot be further reduced. 

However, there is an ample possibility for anthropic models to be
falsified. This could have happened in the case of the cosmological
constant if the observed value turned out to be much smaller than it
actually is. And this may still happen in the future, with improved
understanding of the prior and anthropic factors in the distribution
(\ref{dP}). Also, there is always a possibility that a compelling
non-anthropic explanation for the observed value of $\rV$ will be
discovered. As of today, no such explanation has been found, and the
anthropic model for $\rV$ can certainly be regarded a success. This
may be the first evidence that we have for the existence of a vast
multiverse beyond our horizon.

\section{Acknowlegements}

I am grateful to Jaume Garriga and Ken Olum for comments and to Levon
Pogosian for his help with numerical calculations and the figure.
This work was supported in part by the National Science Foundation.

\end{document}